# On the Four-Dimensional Diluted Ising Model

Giorgio Parisi and Juan J. Ruiz-Lorenzo

Dipartimento di Fisica and Infn, Università di Roma *La Sapienza*

P. A. Moro 2, 00185 Roma (Italy)

parisi@roma1.infn.it ruiz@chimera.roma1.infn.it

**Abstract**

In this letter we show strong numerical evidence that the four dimensional Diluted Ising Model for a large dilution is not described by the Mean Field exponents. These results suggest the existence of a new fixed point with non-gaussian exponents.

# 1 Introduction

Random magnetic systems have been the subject of intensive studies over the last 20 years and much progress has been achieved. The simplest model for a random magnetic system is a ferromagnetic system in which the disorder induces fluctuations in the value of the coupling (or equivalently of the temperature). The simplest realization is a randomly diluted Ising system, where sites (site diluted) or bonds (bond diluted) are randomly removed.

The equivalent Ginsburg Landau model has the following form

$$Z_J = \int \mathrm{d}[\phi] \exp(-S_J[\phi]), \tag{1}$$

where

$$S_J[\phi] = \int \mathrm{d}^D x \left( \frac{1}{2} (\partial_\mu \phi(x))^2 + \frac{1}{2} (m^2 + J(x)) \phi(x)^2 + \frac{g}{4!} \phi(x)^4 \right), \tag{2}$$

and the quenched random variables $J$ are Gaussian distributed with variance

$$\overline{J(x)J(y)} = \lambda \delta(x-y). \tag{3}$$

Here both $\lambda$ and $g$ play the role of coupling constants. It is possible to study analytically this model by considering the case a small coupling constants. In this case perturbation theory may be used to compute the renormalization group flow.

One finds that in four (and more) dimensions the origin is an attractive fixed point, while in less than four dimensions there is a fixed point where both coupling are of order $\epsilon$ in dimensions $D = 4 - \epsilon$. Apart from the detailed problem of computing the fixed point, the situation seems to be clear.

However this result tells us nothing about the possibility of having an other fixed point for large values of the coupling constants. We already know that in the case of a pure system ($\lambda = 0$) there should be no other non trivial fixed points but this statement does not imply that the same scenario is valid for $\lambda$.

Indeed let us suppose to solve the model at fixed non zero $\lambda$ and perform an expansion in $g$. It is extremely difficult to arrive to any conclusion. Indeed one should start by computing the free propagator $G_0(x,y)$, which satisfies the equation

$$(-\Delta + m^2 + J(x)) G_0(x,y) = \delta(x-y), \tag{4}$$

When $m^2$ becomes sufficiently small, $G_0(x,y)$ diverges. In the pure case (i.e. $J=0$) this divergence corresponds to the onset of long range correlations. If we perform a perturbative analysis in $\lambda$, we find that this property holds also at non zero $\lambda$, however a more precise analysis shows that due to non-perturbative effects localized eigenvalues are present.

The transition point is controlled by the extended eigenvalues of the free propagator; therefore also at values of $m^2$ greater than the critical one the quadratic terms has negative eigenvalues and the $g$ expansion is particularly tricky. One may think that the exponent controlling the localization transition are relevant, however they are apparently non trivial also for dimensions greater than 4.

The $g$ expansion at fixed $\lambda$ does seem to lead nowhere. This may leads to the suspicion that there may be two different regimes one for small $\lambda$ and the other for large $\lambda$.

With this motivation we have studied the behavior of a four dimensional diluted spin system, where according to the usual point of view the critical exponents should be those of mean field. We have found that at large dilution the exponent for the susceptibility $\gamma$ is definitely larger that one, thus suggesting that the mean field theory results do not hold. Our simulations have been done for lattices up to $V = 32^4$. We cannot exclude that for larger lattices the behavior of the system crosses over to the mean field, although this possibility is rather unlikely.

## 2 The Model.

The Hamiltonian of the site diluted Ising model can be written in the following form:

$$\mathcal{H} = -\sum_{<i,j>} \epsilon_i S_i \epsilon_j S_j \tag{5}$$

where $< i, j >$ denotes the nearest neighbor pairs, $S_i = \pm 1$ are spin variables and $\epsilon_i$ are independent quenched variables with taking the values 1 and 0 with probability $p$ and $1-p$ respectively, $p$ is the degree of dilution or proportion of spins.

The phase transition disappears for $p$ below a certain value known as $p_c$. We can calculate this value using percolation theory, in four dimensions as $p_c = 0.197$. At this point the critical exponents are $\nu = 0.68$, $\alpha = -0.72$ and $\gamma = 1.44$. It is clear that $\beta_c(p) \to \infty$ when $p \to p_c$, where $\beta_c(p)$ is the critical point of (5) for a given value of dilution [2].

The properties of the model with $p = 1$ are known as it corresponds to the usual Ising Model. There is a second order transition at $\beta_c = 0.1495$ with critical exponents $\alpha = 0$, $\gamma = 1$ and $\nu = 1/2$ (the mean field values)[4].

The influence of dilution on the Ising Model can be studied with the help of the Harris criterion [3, 4]: if the critical exponent $\alpha$ of the undiluted model is greater than zero the critical behavior is modified, otherwise it is not. The present case, in four dimensions, is marginal with $\alpha = 0$ and the criteria does not help us.

Another approach is to use field theoretical methods [1]. If we introduce $n$ replicas we arrive at an $O(n)$ symmetric theory containing a cubic anisotropy term with a coefficient proportional to $1-p$ [3]. By calculating the one loop $\beta-$function of this model and taking the limit $n \to 0$, we find that the only fixed point in four dimensions is Gaussian. Thus, we have the mean field exponents independently of the dilution values [3].

A related model is the random bond Ising Model defined by

$$\mathcal{H} = -\sum_{<i,j>} J_{ij} S_i S_j \tag{6}$$

where the $J_{ij}$ are independent quenched variables taking the values 1 and 0 with probability $p$ and $1-p$ [4].

This model is not identical to the site diluted model because although we can write

$$J_{ij}^{\rm new} \equiv \epsilon_i \epsilon_j \tag{7}$$

these $J_{ij}^{\rm new}$ are not independent. However it is believed that both models are in the same universality class.

# 3 Numerical Method and Observables.

We have used the cluster algorithm due to Wolf [5] for Monte Carlo Simulations. This update method has the advantage that it does not suffer from critical slowing down for the pure model in four dimensions. The dynamical critical exponent for the integrated correlation time of the magnetic susceptibility for the pure model is compatible with zero, $z = -0.10(15)$ [6]. We do not believe that this will be strongly modified in the diluted case. It is easy to translate this algorithm to a diluted Ising model: one simply does not take into account the lattice holes when building a cluster. The average size of clusters is equal to the non-connected magnetic susceptibility for any degree of dilution.

We have measured the non-connected susceptibility ($\chi_{\rm W}$), the total magnetization ($M$), the specific heat ($C$), the Binder cumulant ($B$), the connected susceptibility ($\chi$) and the correlation among the magnetizations of parallel hyperplanes ($G_{\rm plane}(d)$) each defined as follows:

$$\chi_{\rm W} = \frac{1}{V} < M^2 >,$$

$$\chi = \frac{1}{V}(< M^2 > - < |M| >^2),$$

$$C = \frac{1}{V}((< E^2 > - < E >^2),$$

$$B = \frac{1}{2}\left(3 - \frac{\langle M^4 \rangle}{\langle M^2 \rangle^2}\right),$$

$$G_{\rm plane}(d) = \sum_x M(x) M(x+d) \simeq \cosh((d - \frac{L}{2})/\xi) \tag{8}$$

where $V = L^4$ is the volume, $E$ is the total energy, $\xi$ is the correlation length, $M$ is the total magnetization and $M(x)$ is the total magnetization of the hyperplane fixed by $x$. If we label the lattice by $i \equiv (x_1, x_2, x_3, x_4)$ the hyperplane magnetization is

$$M(x_1) = \sum_{x_2, x_3, x_4} S(x_1, x_2, x_3, x_4).$$

If $\beta \ll \beta_c$ we can relate the susceptibilities by

$$\chi = (1 - \frac{2}{\pi})\chi_{\rm W}$$

| $p = 0.8$ | | | | | | |
|---|---|---|---|---|---|---|
| Observable | $\gamma$ | $\beta_c$ | $\chi^2$/d.o.f. | $\gamma$(log) | $\beta_c$ (log) | $\chi^2$/d.o.f.(log) |
| $\chi_W$ $T > T_c$ | 1.13(11) | 0.1894(5) | 0.18 | 1.04(10) | 0.1889(8) | 0.065 |
| $\chi$ $T > T_c$ | 1.17(11) | 0.1895(5) | 0.04 | 1.08(8) | 0.1894(1) | 0.07 |
| $\chi$ $T < T_c$ | 1.11(9) | 0.1894(3) | 2.0 | 1.03(9) | 0.18994(4) | 2 |

Table 1: Fits of the susceptibilities at $p = 0.8$. In the second and third columns we report the results of a pure power fit and in the forth the $\chi^2$ value of the fit. In the last three column the same arrangement but with a power fit with logarithmic dependence as explained in the text.

For completeness we report here the expected critical behavior of the observables:

$$\chi \sim |t|^{-\gamma}$$

$$\xi \sim |t|^{-\nu}$$

$$< m^2 > \sim (-t)^{2\beta} \quad , \quad t < 0 \tag{9}$$

where $\chi$ denotes either $\chi_{\rm W}$ or $\chi$, $t \equiv (T - T_c)/T_c$ is the reduced temperature and $m$ is the intensive magnetization.

To make fits we use the average of the hyperplane-hyperplane correlation functions in the four directions.

We have simulated two different dilutions: $p = 0.8$ and $p = 0.3$. The greater dilution, $p = 0.3$, is not very near to the percolation threshold ($p_c = 0.197$).

We have mainly worked on a large lattice, $24^4$, with periodic boundary conditions and one disorder realization. For the calculations of the correlation length and for some runs at $p = 0.3$ we have used a $V = 32^4$ lattice. With these large lattice sizes we expect that the difference between different realizations of the disorder will be small provided we do not simulate very near to the critical point. We have checked this by comparing the results obtained using different realization of disorder and by matching the $L = 24$ results with the $L = 32$ results. For the results reported in this letter the agreement is very good.

We have run (on WorkStations) 27 different temperatures for the dilution $p = 0.3$ and 22 for $p = 0.8$. A total of five million cluster updates have been done. To estimate the statistical error we have used the jack-knife method.

A source of systematic error is the effect due to the finite size of our lattice. We have used the Binder cumulant to investigate this effect. When the cumulant is different from zero (high temperature phase) or one (low temperature phase) finite size effects are present. Every measurement used in the fits reported in this letter has a Binder cumulant compatible with zero or one. In the thermodynamic limits this parameter tends to the step function with the discontinuity at the transition point.

| $p = 0.8$ | | | |
|---|---|---|---|
| Observable | $2\beta$ | $\beta_c$ | $\chi^2$/d.o.f |
| $< m^2 >$ | 0.82(1) | 0.18939(5) | 0.26 |
| arctanh($< m^2 >$) | 0.84(1) | 0.18935(5) | 0.1 |
| arctanh($< m^2 >$) +log | 0.89(6) | 0.18937(3) | 1.7 |

Table 2: Fits of the magnetization at $p = 0.8$. The same notation as Table 1."+log" denotes a fit with a logarithmic correction as explained in the text .

# 4 Results $p = 0.8$.

We have analyzed the $p = 0.8$ data using (9) and the following Ansatz suggested by the four dimensional $\phi^4$ theory [4] because the $p = 0.8$ dilution is expected to belong in the same universality class as the 4D Ising model and to have the same logarithmic correction:

$$< m >\sim (-t)^\beta (\log(-t))^{1/3} \ , \ \ t < 0.$$

$$\chi \sim \frac{(\log t)^{1/3}}{t^\gamma} \ , \ \ t > 0. \tag{10}$$

In some models arctanh($< m^2 >$) has a better signal than $< m^2 >$, hence we report here the fits of this observable.

We have used the following procedure to find the values of the critical exponents. Firstly we ignore all data with a Binder cumulant different from zero or one. We perform a global fit using the routine MINUIT[7]. We repeat this procedure successfully removing the high temperature data points and monitor the behavior of the effective critical exponent as the data become nearer to the transition point. We observe a plateau and take as our estimate of the critical exponent this plateau.

Our final results for $p = 0.8$ are shown in Table 1 and Table 2. Also we plot in Figure 1(lower) the specific heat against $\beta$.

# 5 Results $p = 0.3$.

With strong dilution $p = 0.3$ we use a pure power fit (9) instead of (10). We analyze the susceptibilities and the correlation length for $T > T_c$. The results for the susceptibilities are reported in Table 3.

To estimate the error on the correlation length we have analyzed the data of the hyperplane-hyperplane correlation with the jack-knife method, estimating for each jack-knife bin the correlation length by means of a $\chi^2$ minimization. Finally we use the jack-knife method again to estimate the error of the previous series of binned correlation lengths. As the mean value we use whose obtaining with the whole set of hyperplane-hyperplane correlations.

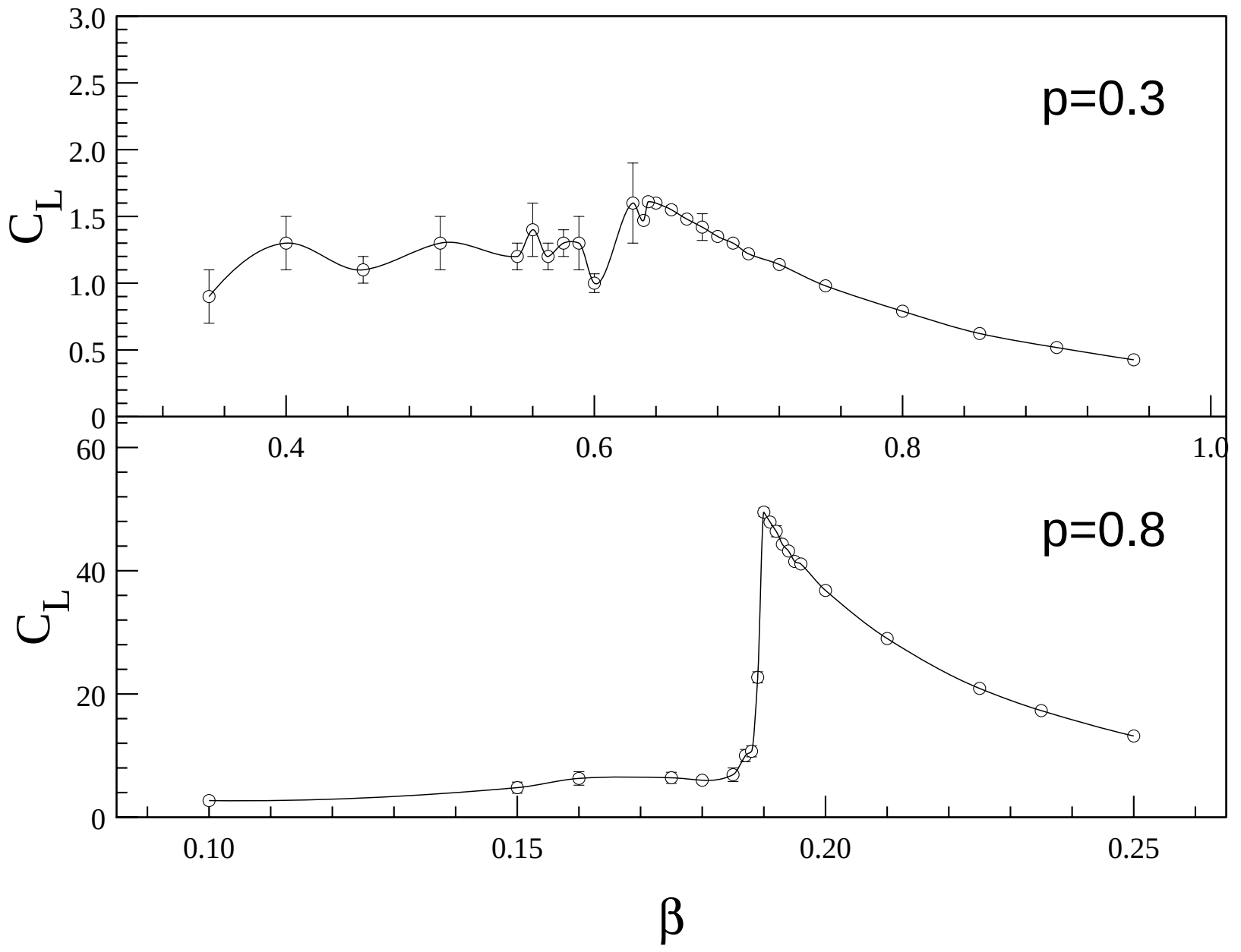

Figure 1: Specific Heat against $\beta$ for the two values of dilutions and $V = 24^4$.

| $p = 0.3$ | | | |
|---|---|---|---|
| Observable | $\gamma$ | $\beta_c$ | $\chi^2$/d.o.f |
| $\chi_{\rm W}$ $T > T_c$ | 1.45(12) | 0.635(4) | 0.9 |
| $\chi$ $T > T_c$ | 1.4(1) | 0.634(4) | 0.50 |

Table 3: Fits of susceptibilities at $p = 0.3$. The same notation of the Table 1, without the log correction in the fit.

Using the $\beta_c$ obtained in the susceptibility fits we calculate the $\nu$ exponent of the correlation length in a two parameter fit, the result is:

$$\xi^{-1} = 2.9(7)[0.635 - \beta]^{0.71(7)} \tag{11}$$

with a $\chi^2/\mathrm{d.o.f} = 0.86$. The largest value of $\xi$ that we have used in the previous fit is $\xi_{\max} = 4.69(5)$. Taking account the error bars on $\beta_c$ in (11) we report the final value:

$$\nu = 0.7(1) \tag{12}$$

In Figure 2 we show the data for the non-connected susceptibility (Fig 2. lower), the connected one (Fig. 2. middle) and the inverse of the correlation length (Fig. 2. upper) along with our best fits for these observables. Also, we plot in Figure 1. the specific heat (Fig 1. upper).

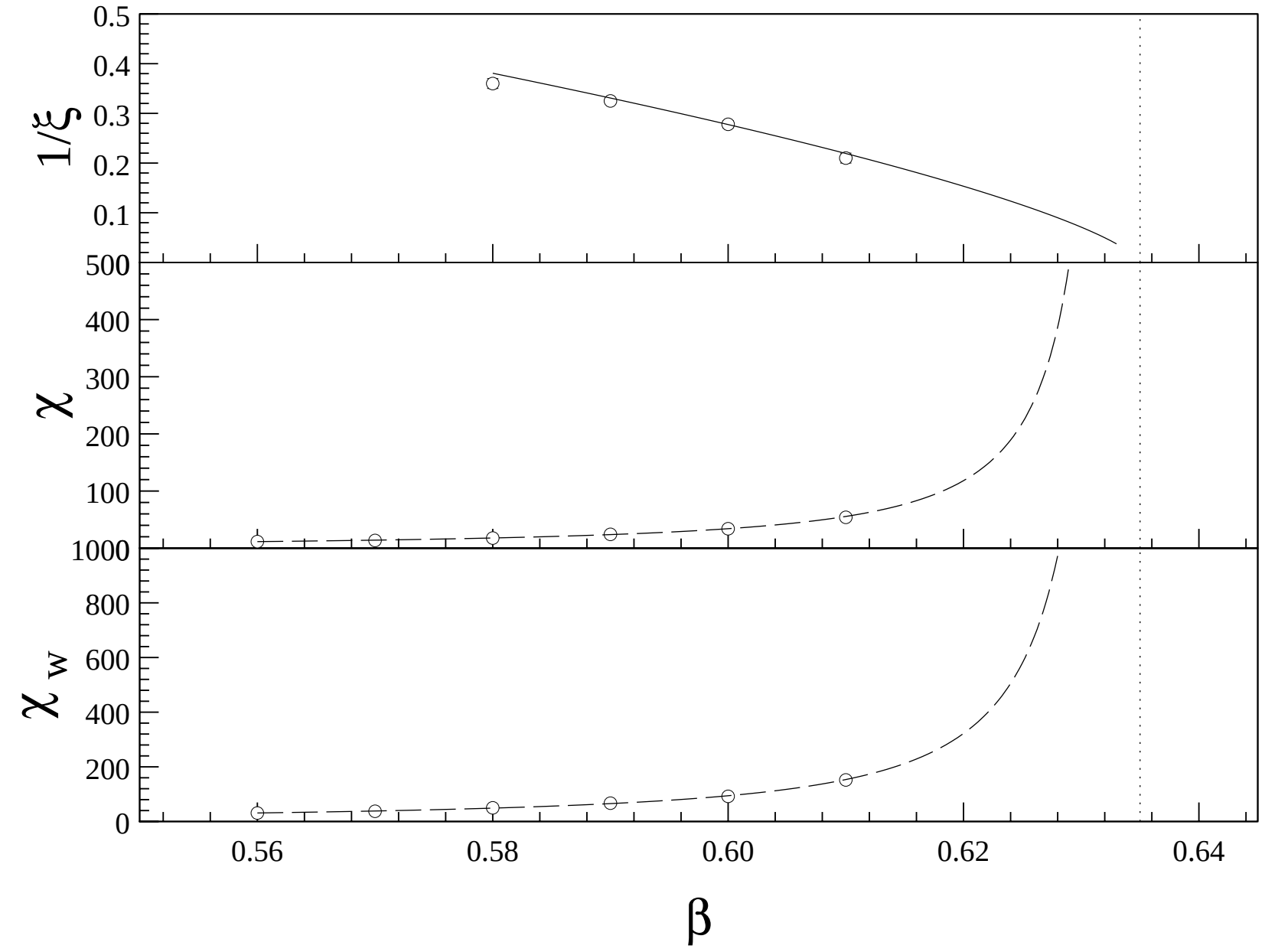

Figure 2: The non-connected susceptibility (lower), the connected one (middle) and the inverse of the correlation length against $\beta$ for $p = 0.3$ and $V = 32^4$. The lines are the fits described in the text. We also mark with a vertical dotted line our best estimate of the critical point.

The specific heat is quite different for the two degrees of dilution. In the case $p = 0.8$ we observe a divergence of this observable while in the case with large dilution the specific heat does not show any divergence. This is already strong indication of the different behavior of the two dilutions.

For $p = 0.8$ we have found critical exponents very similar to those of the pure Ising model.

We have found that the value of the critical exponents show that for lattices up to $V = 32^4$ the system, for $p = 0.3$, is not described by the mean field theory, as one might have believed. Moreover the critical exponents that we have found are very near to those of the pure percolation. A possible explanation would be that the crossover from the percolation to the pure Ising is quite small, however we do not see any indications which point in this direction.

These results suggest the existence of a new fixed point, which can be reached only starting with strong disorder. It would be very interesting to investigate analytically the properties of this fixed point. It may be possible that replica techniques may be useful here.

## Acknowledgments

J. J. Ruiz-Lorenzo is supported by a MEC grant (Spain). It is a great pleasure for us to acknowledge interesting discussions with E. Marinari, G. Harris and D.J. Lancaster.